\documentclass[twocolumn,pre,amssymb]{revtex4}
\usepackage{graphicx}
\usepackage{dcolumn}
\usepackage{bm}
\begin{document}
\title{Density functional theory and demixing of binary hard rod-polymer
mixtures}
\author{P. Bryk}
\affiliation{Department for the Modeling of Physico-Chemical Processes,
Maria Curie-Sk{\l}odowska University, 20-031 Lublin, Poland}
\email{pawel@paco.umcs.lublin.pl}
\date{\today}
\begin{abstract}
A density functional theory for a mixture of hard rods
and polymers modeled as chains built of hard tangent spheres is proposed
by combining the functional due to Yu and Wu for the polymer mixtures
[J. Chem. Phys. {\bf 117}, 2368 (2002)]  with 
the Schmidt's  functional [Phys. Rev. E {\bf 63}, 50201 (2001)]
for rod-sphere mixtures. As a simple application of the functional,
the demixing transition into polymer-rich and rod-rich phases is examined.  
When the chain length increases, the phase boundary broadens
and the critical packing fraction decreases. The shift of the critical
point of a demixing transition is most noticeable for short chains. 
\end{abstract}
\maketitle
The addition of the nonadsorbing polymer to a mono-disperse suspension
of colloidal particles can lead to a phase separation due to
depletion interactions \cite{Asakura54} arising from a tendency
of the system to reduce the volume excluded to the polymer coils.
One of the simplest theoretical models taking into account 
this phenomenon is the so-called Asakura-Oosawa (AO) model of colloid-polymer 
mixtures \cite{Vrij76} where the ideal polymer coils (modeled as spheres)
can freely interpenetrate each other but the polymer-colloid and colloid-colloid 
interactions are of the hard sphere type. Initial studies of such systems focused
on the bulk phase behavior \cite{Gast83,Lekkerkerker92}, however recently developed 
density functional theory (DFT) for the AO model \cite{Schmidt00}
initiated investigations of inhomogeneous colloid-polymer mixtures.
When brought close to a hard wall, such mixtures may develop
a sequence of layering transitions in the partial wetting regime
prior to a transition to complete wetting  \cite{Brader02a,Dijkstra02}.
 
Similar mechanism of fluid-fluid phase separation can be found 
if other mesoscopic particles such as hard rods are used as depletant agents.
Bolhuis and Frenkel (BF) \cite{Bolhuis94} used computer simulations
and free volume theory \cite{Lekkerkerker92}
to study bulk phase behavior of mixtures of colloidal hard spheres and vanishingly thin hard rods.
They found a surprisingly good agreement (c.f. Fig.~3 in \cite{Bolhuis94})
between the free volume theory and Gibbs ensemble Monte Carlo simulations.
Although the vanishing thickness and volume constitutes a significant simplification
(e.g. it rules out the isotropic-nematic transition),
the BF model is relevant for some experimental systems e.g. mixtures of silica spheres
and silica coated bohemite rods 
\cite{Koenderink99,Vliegenthart99}. In order to describe inhomogeneous 
hard rod-hard sphere systems Schmidt \cite{Schmidt01} proposed a density functional
theory which in its structure closely resembles the AO functional \cite{Schmidt00}.
The functional for the BF model incorporates the exact low-density limit and
yields the same equation of state as in Ref.~\cite{Bolhuis94}.
Moreover, entropic surface phase transitions 
found previously in model colloid-polymer mixtures close to a hard wall
were also recently encountered in hard rod-hard sphere mixtures
\cite{Roth03}. This further demonstrates deep similarities between the two models.

The aim of the present work is to construct a functional for a mixture of
vanishingly thin hard rods and polymers.
Such systems can be regarded as simple microscopic models of the liquid crystal-polymer mixtures. 
The functional is constructed by combining the Schmidt's functional for the BF model 
with the Yu and Wu \cite{Yu02} functional for mixtures of polymeric fluids.
To make this conjecture we take the advantage of the fact that both functionals
underlie the fundamental measure theory (FMT) of Rosenfeld \cite{Rosenfeld89}.
As a simple application we investigate bulk phase diagrams resulting from the proposed theory.

Consider a mixture of hard, vanishingly thin needles (species $N$) 
of length $L$ and polymers (species $P$) modeled as chains composed
from $M$ tangentially bonded hard-sphere segments of diameter $\sigma$. 
The hard-sphere monomers building up the chains are freely jointed i.e. 
they can adopt any configuration as long as it is free of the 
intermolecular and intramolecular overlap.
The interaction potential between needles $V_{NN}=0$ for all separations,
while the pair potential between a polymer segment and a hard rod, $V_{PN}$, 
and between two polymer segments, $V_{PP}$, is of a hard-core type i.e. 
is infinite if a pair of objects overlap and zero otherwise.
The grand potential of such system as a
functional of local densities of polymers
$\rho_{P}({\bf R})$ and needles $\rho_{N}({\bf r},{\bm \omega})$
can be written as
\begin{eqnarray}
\lefteqn{\Omega[\rho_P({\bf R}),\rho_{N}({\bf r},{\bm \omega})]=
F_{id}[\rho_P({\bf R}),\rho_{N}({\bf r},{\bm \omega})]+}\nonumber\\
& & F_{ex}[\rho_P({\bf R}),\rho_{N}({\bf r},{\bm \omega})]+
\int\!\!d{\bf R}\rho_{P}({\bf R})(V_{ext}^{(P)}({\bf R})-\mu_{P})\nonumber\\
&+&\int\!\!d{\bf r}\int\!\frac{d{\bm \omega}}{4\pi}\rho_{N}({\bf r},{\bm \omega})
(V_{ext}^{(N)}({\bf r},{\bm \omega})-\mu_{N})\;,
\end{eqnarray}
where ${\bf R}\equiv ({\bf r}_{1}, {\bf r}_{2}, \cdots, {\bf r}_{M})$ denotes
a set of coordinates describing
the segment positions, ${\bm \omega}$ describes the orientation of the rod,
$V_{ext}^{(P)}({\bf R})$, $\mu_P$,  $V_{ext}^{(N)}({\bf r},{\bm \omega})$
and $\mu_N$ are the external and the chemical potentials for
polymers and rods, respectively.
The ideal part of the free energy is known exactly 
\begin{eqnarray}
\lefteqn{\beta F_{id}[\rho_P({\bf R}),\rho_{N}({\bf r},{\bm \omega})]=
\beta\int\!\!d{\bf R}\rho_{P}({\bf R})V_{b}({\bf R})+}\nonumber\\
& &\int\!\!d{\bf R}\rho_{P}({\bf R})
[\ln(\Lambda^{3}_{P}\rho_{P}({\bf R}))-1]+\nonumber\\
& &
\int\!\!d{\bf r}\int\!\frac{d{\bm \omega}}{4\pi} \rho_{N}({\bf r},{\bm \omega})
[\ln(\Lambda^{3}_{N}\rho_{N}({\bf r},{\bm \omega}))-1]\;,
\end{eqnarray}
where $\Lambda_P$ and $\Lambda_N$ are the thermal wavelengths for 
polymers and needles,
respectively. The total bonding potential $V_{b}({\bf R})$ is a sum
of bonding potentials $v_b$ between the segments 
$V_{b}({\bf R})=\sum_{i=1}^{M-1}v_b(|{\bf r}_{i+1}-{\bf r}_{i}|)$
and for the tangential hard spheres this contribution satisfies
$\exp [-\beta V_{b}({\bf R})]=
\prod_{i=1}^{M-1}\delta(|{\bf r}_{i+1}-{\bf r}_{i}|-\sigma)/4\pi\sigma^{2}$.

Within the framework of the fundamental measure theory 
the excess free energy density $\Phi$ is expressed as
a simple {\it function} of the weighted densities $n_{\alpha}^{(i)}$.
We assume that $\Phi$ can be represented as a sum of the orientation-independent
polymer contribution, $\Phi_{POL}$, and the orientation-dependent polymer-needle
contribution, $\Phi_{PN}$.
For the orientation-independent contribution we use
an extension of the Rosenfeld's FMT to the polymeric fluids \cite{Yu02},
where the polymer excess free energy density $\Phi_{POL}$ is assumed 
to be a functional of only segment
densities $\rho_{PS}({\bf r})$ defined as
\begin{equation}
\rho_{PS}({\bf r})=\sum_{i=1}^{M}\rho_{PS,i}({\bf r})=\sum_{i=1}^{M}
\int\!\!d{\bf R}\delta({\bf r}-{\bf r}_i)\rho_{P}({\bf R})\;,
\end{equation}
where $\rho_{PS,i}({\bf r})$ is the local density of the segment $i$
of the polymer. Following \cite{Yu02} we assume that $\Phi_{POL}$
can be split into the hard-sphere $\Phi_{HS}$ contribution 
resulting from the hard-sphere repulsion of polymer segments
and $\Phi_{P}$ contribution arising due to the chain connectivity.
Taking into account the assumptions mentioned above
we have $\Phi=\Phi_{PN}+\Phi_{HS}+\Phi_{P}$.
The total excess free energy of the inhomogeneous system is obtained via the
integration of the excess free energy density
\begin{equation}
\beta\!F_{ex}\!=\!\int\!\!d{\bf r}\!\!\int\!\frac{d{\bm \omega}}{4\pi}
\Phi_{PN}(\{n_{\alpha}^{(i)}\})+\Phi_{HS}(\{n_{\alpha}^{(P)}\})+
\Phi_{P}(\{n_{\alpha}^{(P)}\}).
\end{equation}

For the hard-sphere part $\Phi_{HS}$ we choose the original Rosenfeld 
expression \cite{Rosenfeld89}
\begin{eqnarray}\label{eq:5}
\Phi_{HS}(\{n_{\alpha}^{(P)}\})&=&
-n_0^{(P)} \ln (1-n_{3}^{(P)})+\\
\frac{n_{1}^{(P)}n_{2}^{(P)}-
{\bm n}_{V1}^{(P)}\cdot {\bm n}_{V2}^{(P)}}{1-n_{3}^{(P)}}
&+&\frac{(n_{2}^{(P)})^{3}-3n_{2}^{(P)}{\bm n}_{V2}^{(P)}
\cdot{\bm n}_{V2}^{(P)}}
{24\pi (1-n_{3}^{(P)})^{2}}\nonumber\;.
\end{eqnarray}
The polymer weighted densities $n_{\alpha}^{(P)}({\bf r})$ 
are evaluated via spatial convolutions
\begin{equation}\label{eq:6}
n_{\alpha}^{(P)}({\bf r})=\int\!\!d{\bf r}' \rho_{PS}({\bf r}')w_{\alpha}^{(P)}({\bf r}-{\bf r}') \;,
\end{equation}
where the weight functions  $w_{\alpha}^{(P)}({\bf r})$ are given by
\begin{eqnarray}
w^{(P)}_{3}({\bf r})=\Theta(\frac{\sigma}{2}-|{\bf r}|)\;&,&\;
w_{2}^{(P)}({\bf r})=\delta (\frac{\sigma}{2}-|{\bf r}|)\;,\\
{\bm w}_{V2}^{(P)}({\bf r})=\frac{{\bf r}}{|{\bf r}|}\delta (\frac{\sigma}{2}-|{\bf r}|)\;
&,& w_{1}^{(P)}({\bf r})=\frac{w_{2}^{(P)}({\bf r})}{2\pi\sigma}\;,\\
w_{0}^{(P)}({\bf r})=\frac{w_{2}^{(P)}({\bf r})}{\pi\sigma^2}\;
&,&{\bm w}_{V1}^{(P)}({\bf r})=\frac{{\bm w}_{V2}^{(P)}({\bf r})}{2\pi\sigma}\;.
\end{eqnarray}
The contribution due to the chain connectivity is evaluated using
the Wertheim's first-order perturbation theory for a bulk fluid \cite{Wertheim87} and extended 
(using FMT-style weighted densities)
by Yu and Wu to inhomogeneous systems \cite{Yu02}
\begin{equation}\label{eq:10}
\Phi_{P}(\{n_{\alpha}^{(P)}\})=\frac{1-M}{M}n_0^P\zeta\ln[y_{hs}(\sigma,\{n_{\alpha}^{(P)}\})]\;,
\end{equation}
where $\zeta=1-{\bm n}_{V2}^{(P)}\cdot{\bm n}_{V2}^{(P)}
/(n_2^{(P)})^2$, while $y_{hs}$ is connected with the
Carnahan-Starling expression for the contact value of the radial distribution function
of a hard-sphere mixture
\begin{equation}\label{eq:11}
y_{hs}(\sigma,\{n_{\alpha}^{(P)}\})=\frac{1}{1-n_3^{(P)}}+
\frac{n_2^{(P)}\sigma\zeta}
{4(1-n_3^{(P)})^2}+\frac{(n_2^{(P)})^2\sigma^2\zeta}{72(1-n_3^{(P)})^3}
\end{equation}
We note here that both $\Phi_{HS}$ and $\Phi_{P}$ are 
independent on the local  density of rods.

To specify the polymer-needle contribution we make use of Schmidt's DFT
for hard rod-sphere mixtures \cite{Schmidt01} and write
the excess free energy density due to 
vanishingly thin needles as
\begin{equation}\label{eq:12}
\Phi_{PN}(\{n_{\alpha}^{(i)}\})=
-n_{0}^{(N)}\ln(1-n_3^{(P)})+\frac{n_1^{(N)} n_2^{(PN)}}{1-n_3^{(P)}}\;.
\end{equation}
The equation above implies that the polymer-needle contribution 
to the excess free energy
stems solely from the hard-core repulsion between the needle
and the hard sphere forming the polymer segment. 

The needle weighted densities, $n_{\alpha}^{(N)}$, are obtained through spatial 
convolutions of the needle local density and the corresponding 
orientation-dependent weight functions
\begin{equation}\label{eq:13}
n_{\alpha}^{(N)}({\bf r},{\bm \omega})=
\int\!\!d{\bf r}'\;\rho_{N}({\bf r}',{\bm \omega})
w_{\alpha}^{(N)}({\bf r}-{\bf r}',{\bm \omega}) \;,\;\; \alpha=0,1,
\end{equation}
while the ``mixed'' polymer segment-needle weighted density,
$n_{2}^{(PN)}$, is obtained via spatial convolution of the polymer 
segment density and an orientation-dependent weight function
\begin{equation}\label{eq:14}
n_{2}^{(PN)}({\bf r},{\bm \omega})=\int\!\!d{\bf r}'\;\rho_{PS}({\bf r}')
w_{2}^{(PN)}({\bf r}-{\bf r}',{\bm \omega}) \;.
\end{equation}
Corresponding weight functions are given as \cite{Schmidt01}
\begin{eqnarray}
w^{(N)}_{1}({\bf r},{\bm \omega})&=&
\frac{1}{4}\int_{-L/2}^{L/2} dl \delta({\bf r}+{\bm \omega} l)\;,\nonumber\\
w^{(N)}_{0}({\bf r},{\bm \omega})&=&
\frac{1}{2}[\delta({\bf r}+{\bm \omega} l)+
\delta({\bf r}-{\bm \omega} l)]\;,\nonumber\\
w^{(PN)}_{2}({\bf r},{\bm \omega})&=&2|{\bm w}_{V2}^{(P)}({\bf r})
\cdot{\bm \omega}|\;.
\end{eqnarray}
This completes the prescription for the functional.
The present theory  reduces to the Schmidt's functional
\cite{Schmidt01} if $M=1$ and to Yu and Wu functional \cite{Yu02} 
if the density of rods $\rho_N=0$. 
Note also that, similarly to \cite{Schmidt01},
the functional is {\it linear} in the local density of rods.
\begin{figure}
\includegraphics[clip,width=8cm]{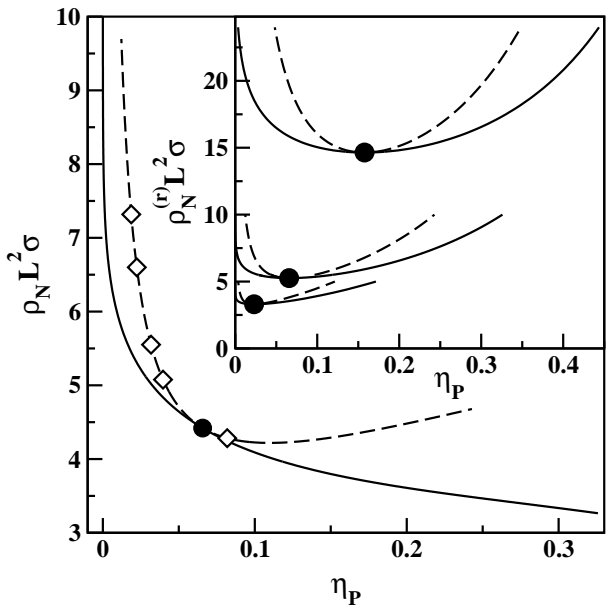}
\caption{\label{fig:1}  Phase diagrams for hard rod-polymer mixtures.
Solid and dashed lines denote the binodals and spinodals,
respectively. Black circles indicate the critical points of a demixing 
transition. The main plot shows phase diagram for a mixture of
hard rods with size ratio $q=L/\sigma=1$
and hard-sphere 10-mers, in terms of polymer packing fraction 
$\eta_P=\frac{\pi}{6}\sigma^{3}M\rho_P$ and the scaled needle density
$\rho_NL^2\sigma$. Open diamonds denote the
critical points for a mixture of hard-sphere 10-mers and hard rods
with size ratios $q=20, 15, 8, 5$ and 0.1 (from top to bottom).
The inset shows the polymer packing fraction-scaled needle reservoir
density representation of the phase diagrams for a mixture of hard rods 
with size ratio $q=1$ and polymers with $M=1,10$ and 100 (from top to bottom).
}
\end{figure}

Although the proposed DFT is intrinsically designed to study inhomogeneous systems,
here we restrict ourselves to the isotropic bulk phases where the density of both species 
are constant. In this case the scalar weighted densities become proportional
to the bulk densities of needles and polymers, while the vector weighted densities
vanish. Consequently from Eq.~\ref{eq:6} we obtain 
$n_{\alpha}^{(P)}=\xi_{\alpha}^{(P)} M\rho_{P}=\xi_{\alpha}^{(P)}\rho_{PS}$, 
with $\xi_3^{(P)}=\pi/6\,\sigma^3$, $\xi_2^{(P)}=\pi\sigma^2$,
$\xi_1^{(P)}=\sigma/2$ and $\xi_0^{(P)}=1$. Likewise Eq.~\ref{eq:13} yields
$n_{\alpha}^{(N)}=\xi_{\alpha}^{(N)} \rho_{N}$ where $\xi_1^{(N)}=L/4$
and $\xi_0^{(N)}=1$. Finally Eq.~\ref{eq:14} leads to
$n_2^{(PN)}=\xi_2^{(P)} \rho_{PS}$. 
After inserting the above expressions for the weighted densities into
the Eq.~\ref{eq:5} and Eqs.~\ref{eq:10}-\ref{eq:12}  
the total free energy per unit volume, $\Phi_v$,
is evaluated as $\Phi_v=\Phi_{HS}+\Phi_{P}+\Phi_{PN}+
\beta^{-1}\rho_{P}[\ln(\rho_{P}\Lambda^3_P)-1]+
\beta^{-1}\rho_{N}[\ln(\rho_{N}\Lambda^3_N)-1]$.
From $\Phi_v$ the pressure, $p$, and the chemical potentials of both species are
easily calculated
\begin{equation}
\beta p=-\Phi_v+\sum_{j=P,N}\rho_j\frac{\partial\Phi_v}
{\partial \rho_j}\;,\;\;\beta\mu_j=\frac{\partial\Phi_v}{\partial\rho_j}\;.
\end{equation}
Under appropriate conditions a mixture of polymers and hard rods
undergoes entropically driven demixing transition to polymer-rich
(rod-poor) and polymer-poor (rod-rich) phases.
The coexsisting equilibrium densities (binodals)  
were obtained by solving simultaneously equations for the equality 
of pressures and chemical potentials in two phases.
The spinodals delimiting the regions stable against fluctuations
of density and composition were evaluated from the condition
$\det[\partial^2\Phi_v/\partial\rho_i\partial\rho_j]=0$, $i,j=P,N$.
The critical points were evaluated from
\begin{eqnarray}\label{eq:17}
s^3\frac{\partial^3\Phi_v}{\partial\rho_P^3}+3s^2
\frac{\partial^3\Phi_v}{\partial\rho_P^2\partial\rho_N}&+&
3s\frac{\partial^3\Phi_v}{\partial\rho_P\partial\rho_N^2}+
\frac{\partial^3\Phi_v}{\partial\rho_N^3}=0\,,\nonumber\\
s\equiv\frac{-\partial^2\Phi_v}{\partial\rho_P\partial\rho_N}&/&
\frac{\partial^2\Phi_v}{\partial\rho_P^2}\;.
\end{eqnarray}
The above equation arises from the fact that at the critical point
the tie-line connecting the coexisting densities
becomes tangent to the spinodal line \cite{Rowlinson59}.

In Fig.~\ref{fig:1} we show examples of binodals (solid lines),
spinodals (dashed lines) and critical points (black circles)
resulting from the present theory.
The phase diagrams plotted in the polymer packing fraction,
$\eta_P=\frac{\pi}{6}\sigma^{3}M\rho_P$
versus scaled needle reservoir density, $\rho_N^{(r)}L^2\sigma$
representation (see the inset)
were evaluated for systems with $M=1,10,100$ for constant $q=L/\sigma=1$. 
For the special case $M=1$ (the uppermost diagram) the system reduces
to the BF model \cite{note1}.

As the chain length is increased, the phase boundary becomes more asymmetric
and the critical point moves towards lower 
polymer packing fractions and towards lower reservoir needle densities.
Another interesting feature observed in \cite{Schmidt01} is that 
when the polymer packing fraction-{\it actual} needle density representation
instead of the polymer packing fraction-{\it reservoir} needle density
representation is chosen, the critical points for different size
ratios $q$ are located on the spinodal of the system with
$q=1$. We find the same behavior also in the present model
i.e. for $M>1$. In the main plot of Fig.~\ref{fig:1} we show
phase diagram for a mixture of 10-mers and hard rods.
The critical points for mixtures with
different size ratios (open diamonds, from top to bottom for 
$q=20$, 15, 8, 5  and 0.1) lie on the spinodal for the system
with $q=1$ (dashed line).
\begin{figure}[t]
\includegraphics[clip,width=8cm]{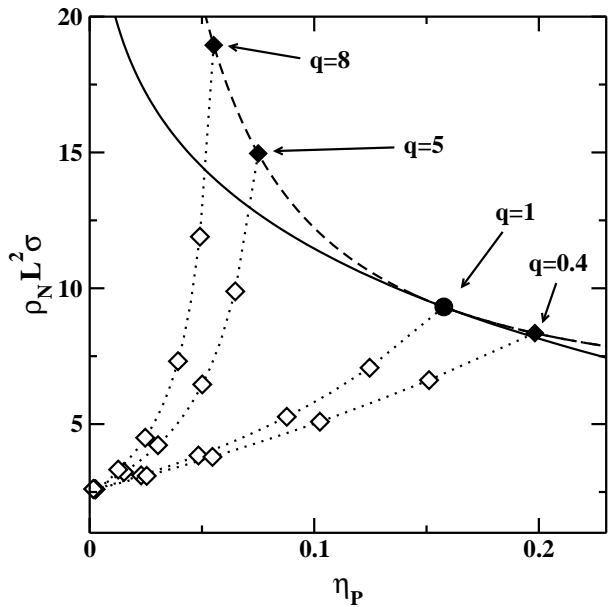}
\caption{\label{fig:2}
The phase diagram for a hard rod-hard sphere monomer ($M=1$)
mixture with the size ratio $q=1$. Solid and dashed line
denote the binodal and the spinodal, respectively.
Black diamonds on the spinodal indicate the critical points
of a demixing transition for monomer-hard rod mixtures with 
different size ratios $q=8$, 5, 1 and 0.4 (going from top to bottom).
Open diamonds denote the critical points of a demixing transition
for a mixture of polymers consisting of $M$ monomers and hard rods
for fixed $q$ and for $M=2$, 5, 20, 100 and 10000
(starting from the right hand side).
Dotted lines serve as a guide to the eye and connect the
critical points for the systems with the same size ratio $q$.}
\end{figure}

Let us now consider the limit of very long polymer chains.
Intuitively one could argue, that in this regime the 
phase behavior of a needle-polymer mixture should barely depend on 
the rod length to polymer-segment diameter ratio $q=L/\sigma$ because 
the physical dimensions of the polymer e.g. the gyration radius,
$R_g$ become much bigger than the rod elongation.
This scenario is captured
within the present theory. In Fig.~\ref{fig:2} we show the phase diagram 
for a mixture of hard-sphere monomers ($M=1$) and hard rods for 
the size ratio $q=1$.
The critical points for different size ratios $q$ (black diamonds)
are located on the spinodal. Open diamonds denote the critical points 
for the systems with $M=2$, 5, 20, 100, and 10000, respectively.
As the chain length increases the critical points for the systems
with different size ratios but the same $M$ come closer together
and for $M=10000$ they virtually merge (on the figure scale) into one point.
In the limit $M\to\infty$ the critical scaled needle density
tends to the value 2.5526 for all $q$ while 
the critical polymer packing fraction tends to zero.
Similar limiting behavior was found in mixtures of spherical
colloids and polymers with excluded volume interactions 
\cite{Bolhuis02,Sear02,Paricaud03}. 

In conclusion, in this work we propose a density functional theory
for a mixture of vanishingly thin hard rods and polymers modeled
as chains built of hard tangent spheres. The functional is constructed
by combining the functional due to Yu and Wu for polymer mixtures \cite{Yu02}
with Schmidt's functional for hard rod-sphere mixtures \cite{Schmidt01}.
The proposed theory predicts a demixing transition similar in its nature
to that observed for sphere-rod systems. The present functional is
well suited to study inhomogeneous systems. It would be of interest
to consider a fluid-fluid interface or to investigate
surface phase transitions such as entropic wetting or layering that
have been discovered in colloid-polymer and colloid-rod mixtures.
It is also straightforward to incorporate the Onsager limit \cite{Brader02b}
of the needle contribution to the functional generating thus
a more sophisticated model of inhomogeneous liquid crystal-polymer mixtures.
Some of these topics are already under study in our laboratory.

This work has been supported by KBN of Poland under the Grant
"Wp{\l}yw samoorganizacji na r\'ownowagi fazowe w p{\l}ynach z{\l}o{\.z}onych".

\end{document}